\documentclass[twocolumn,showpacs,preprintnumbers,amsmath,amssymb,nofootinbib]{revtex4-2}

\usepackage{marvosym}
\usepackage{comment} 
\usepackage{cancel}
\usepackage{graphicx}
\usepackage{dcolumn}
\usepackage{bm}
\usepackage[compact]{titlesec}

\makeatletter
\DeclareRobustCommand{\cev}[1]{%
  {\mathpalette\do@cev{#1}}%
}
\newcommand{\do@cev}[2]{%
  \vbox{\offinterlineskip
    \sbox\z@{$\m@th#1 x$}%
    \ialign{##\cr
      \hidewidth\reflectbox{$\m@th#1\vec{}\mkern4mu$}\hidewidth\cr
      \noalign{\kern-\ht\z@}
      $\m@th#1#2$\cr
    }%
  }%
}
\makeatother
\usepackage{xcolor}

\usepackage{hyperref}
\renewcommand{\arraystretch}{2.7}       
\newcommand{\Gini}{{\small{\mathsf{G}}}}

\newcommand{\mbf}{\mathrm{{b/f}}}
\newcommand{\mb}{\mathrm{b}}
\newcommand{\mf}{\mathrm{f}}
\newcommand{\mP}{\mathrm{p}}

\newcommand{\be}{\begin{equation}}
\newcommand{\ee}{\end{equation} }
\newcommand{\beqa}{\begin{eqnarray} }
\newcommand{\eeqa}{\end{eqnarray} }
\newcommand{\ba}{\begin{array}}
\newcommand{\ea}{\end{array}}
\newcommand{\bpm}{\begin{pmatrix}}
\newcommand{\epm}{\end{pmatrix}}

\newcommand{\dis}{\displaystyle}

\newcommand{\rmd}{{\rm d}}
\newcommand{\rd}{{\rmd}}

\newcommand\cN{{\cal N}}

\newcommand\cP{{\cal P}}

\def\brd{\bar{d}}

\def\bri{\bar{\imath}}

\def\brbeta{\bar{\beta}}

\def\brm{{\bar{m}}}

\def\brM{\bar{M}_{w}}

\def\brP{\bar{P}}

\def\brZ{\bar{Z}}

\begin{document}

\title{Distinguishable Cash,  Bosonic  Bitcoin, and Fermionic   Non-fungible Token}

\author{Zae Young Kim}
\affiliation{Center for Quantum Spacetime, Sogang University, 35 Baekbeom-ro, Mapo-gu, Seoul 04107,  Korea}

\author{Jeong-Hyuck Park}
\email{park@sogang.ac.kr}
\affiliation{Department of Physics, Sogang University, 35 Baekbeom-ro, Mapo-gu, Seoul 04107,  Korea}

\begin{abstract}
\noindent{Modern technology has brought novel types of wealth. In contrast to hard cash, digital currency  does not have a physical form. It exists in electronic forms only.  To date, it has not been clear what impacts its ongoing growth will have, if any, on wealth distribution. Here, we propose to identify all forms of contemporary wealth into two classes: \textit{distinguishable} or \textit{identical}. Traditional tangible moneys are all distinguishable. Financial assets and cryptocurrencies, such as bank deposits and Bitcoin, are boson-like, while non-fungible tokens are fermion-like. We derive their ownership-based  distributions in a unified manner.  Each class follows essentially the Poisson or the geometric distribution. We contrast their distinct  features such as   Gini coefficients. Furthermore, aggregating different kinds of wealth corresponds to a weighted convolution where the number of banks matters and Bitcoin follows Bose--Einstein distribution. {Our proposal opens a new avenue to understand the  deepened  inequality in modern economy, which is based on the statistical  physics property of wealth  rather than the individual ability of owners. We call for verifications with real data.}
}
\end{abstract}

                             
\maketitle

\textit{Introduction.}---When two one-dollar banknotes  are randomly gifted   to two people,  there occur total four possible ways of distributions.  While counting so,   it has been naturally assumed that both notes are \textit{distinguishable} from each other, since they are  for sure distinct physical objects, not to mention the different serial numbers printed on them.   In contrast, when two cents are credited to a pair of savings bank accounts,  there are  three possibilities,  because the two cents as deposits   are indistinguishable. Deposits do not have  a physical form. They  exist in the form of abstract numbers by `claim' and `trust' between  the bank and the account holders.     
While \textit{one}'s can  add up to a natural number, say $k\in\mathbb{N}$, 
\be
1+1+\cdots +1=k\,,
\label{Natural}
\ee 
all the \textit{one}'s are intrinsically \textit{identical} and indistinguishable from one another.  The notion of being indistinguishable, or interchangeably {identical},  is a fundamental property of elementary particles  in  physics:  bosons can share quantum states but fermions subject to the Pauli exclusion principle cannot. Consequently, their  statistical distributions differ significantly.    While the identical property holds certainly for  particles  at  quantum scale, there appears no clear-cut limit of applicability to larger macroscopic  objects.

In this paper, we propose to identify  all kinds of wealth  into two classes:  \textit{distinguishable} or  \textit{identical}.   All the traditional  tangible moneys \textit{i.e.~}hard cash including minted coins and banknotes are of physical existence and  belong to the   distinguishable class.  In contrast,  financial assets   like  bank deposits, stocks, bonds, and loans belong to the boson-like identical   class.  Furthermore, all the electronic forms of wealth  share the identical property.   At deep down level of  information technology or    atomic physics, they comprise of    chain of  \textit{bits} which have finite length. The pieces of information stored are accordingly  limited  mostly to the amounts   and, hence, are  abstract like the deposit or the natural number~(\ref{Natural}).   With no restriction on the amount of  possession,  cryptocurrencies, \textit{e.g.~}Bitcoin~\cite{Bitcoin}  are  boson-like.   Contrarily, having unique digital identifiers,  non-fungible tokens (NFTs) may be identified as fermions.  Having said so,   we shall  demonstrate that generic  identical  wealth can be universally and effectively described by  Gentile statistics~\cite{Gentile}  which postulates  a cutoff for the maximal amount of possession.

It is an established  fact that distinguishable, bosonic, and fermionic particles follow respectively the  Maxwell--Boltzmann, Bose--Einstein, and Fermi--Dirac statistics, which are all about  the number of the particles themselves for a given energy.  On the contrary,  our primary interest in this work is to  derive the ownership-based  distributions of  wealth, \textit{i.e.~}the number of owners who possess a certain  amount of wealth, while the owners are  assumed to  be   always  distinguishable.      {Further, it is our working assumption that  wealth is  distributed in a `random' manner. This should be the case  if ideally  the owners were  all   equal.}  {It goes beyond the scope of the present paper to test the hypothesis against real data.}

\textit{Basic scheme through  elemental examples.}---We start with  an elementary example  of distributing   $M$ number of minted one-cent coins  to  $N$ number of people in a random manner.  We let $n_{k}$ be the  number of people each of whom owns $k$ number of coins,  $k=0,1,2,\cdots$.   As we focus on `private ownership' meaning no allowance of sharing, the opposite notion ``$k_{n}$'' does not make sense (except $k_{n=1}$), which in a way breaks the symmetry between people and coins both of which are distinguishable.  There are two constraints $n_{k}$'s  satisfy 
\be
\ba{ll}
{\sum_{k=0}^{\infty}~n_{k}=N\,,}\qquad&\qquad
{\sum_{k=0}^{\infty}~kn_{k}=M\,.}
\ea
\label{constraints0}
\ee
Irrespective of  our notation,   an effective upper bound in the sums exists such as  $0\leq k\leq M$.  Our primary aim is to compute the total number of all  possible or  `degenerate'  ways of distributions for a given  set $\{n_{k}$'s$\}$.  Hereafter, generically for any kinds of wealth, we denote such a total number by $\Omega$ and further  factorise it into two numbers, $\Omega=\Upsilon\times\Phi$, where  $\Upsilon$ is all about the grouping of the owners into $\{n_{k}$'s$\}$ and thus  is    independent of the sorts of wealth.  The  properties of  wealth are to be reflected in $\Phi$. Specifically,   the  total number of possible cases for  the $N$ number of people to be grouped into $n_{0},n_{1},n_{2},\cdots$ is  
\be
\Upsilon=
\frac{N!}{n_{0}!n_{1}!n_{2}!\cdots}=\frac{N!}{\prod_{k=0}^{\infty}n_{k}!}\,.
\label{Upsilon0}
\ee
While so, that  for the  $M$ coins  to be grouped into 
\be
\underbrace{1,1,\cdots,1}_{n_{1}},\,\underbrace{2,2,\cdots,2}_{n_{2}},\,\cdots\cdots\,\underbrace{k,k,\cdots,k}_{n_{k}},\cdots\,,
\label{ngroup}
\ee
is, as the coins are distinguishable, 
\be
\Phi=
\frac{M!}{(1!)^{n_{1}}(2!)^{n_{2}}\cdots}=\frac{M!}{\prod_{k=1}^{\infty}(k!)^{n_{k}}}\,.
\label{Phi0}
\ee
Crucially, for each  case  in $\Upsilon$,  any of $\Phi$ can equally occur. Thus, the total number of possible  distributions for a given set $\{n_{k}$'s$\}$ is    the product $\Upsilon\Phi=\Omega$.    The degeneracy $\Phi$ as counted in (\ref{Phi0}) is  significant since it depends on $n_{k}$'s. Insignificant degeneracies  that are independent of $n_{k}$'s may be taken into account which will multiply $\Phi$  by an overall constant.  For example,  extra  distinctions  depending on whether the distribution of each coin occurs in the morning or afternoon will give an overall factor $2^{M}$ to $\Phi$.  Yet, our primary interest is to obtain the most probable  distribution of $n_{k}$.  Following  the standard analysis in statistical physics at equilibrium,  \textit{e.g.~}\cite{Huang},  we shall assume $N$ to be sufficiently large,  apply  the variational method induced by $\delta n_{k}$  to $\ln\Omega=\ln\Upsilon+\ln\Phi$,  and  acquire the extremal solution.   Accordingly, any insignificant degeneracy independent of $n_{k}$'s  becomes  irrelevant and ignorable. It merely shifts $\ln\Phi$ by a constant.

We    turn to  savings accounts. We consider  the $M$ cents to be now credited to distinguishable $N$ savings accounts. Since deposits are boson-like identical,  the total number of possible distributions $\Omega$ is essentially  $\Upsilon$~(\ref{Upsilon0}) itself up to  multiplying  an insignificant overall constant. This irrelevant degeneracy can arise when the bank accounts keep records of all the details of the crediting of the deposits, 
\textit{e.g.~}the time of transaction,  which would make  the credited $M$ cents  to appear  seemingly   distinguishable.  However, all the information of each credit are recorded in a chain of bits which has a finite length, say  $l=l_{0}+l_{1}$ that decomposes into  $l_{0}$ for the very record of the amount   $k$ and  $l_{1}$ reserved for any extra information. While the former  is rigidly fixed, the extra pieces of information are rather stochastic  and hence contribute to $\ln\Phi$ by a constant shift, ${l_{1}\!}\ln 2$, which is hence ignorable.\footnote{In this reason, we prefer to say  credits are boson-like rather than (precisely) bosons.   Further, we note that the extra information is generically   postdictive: they do not preexist before the transactions take place, or before the ownerships settle down.}  

Lastly, fermion-like wealth or NFTs set  $M=1$ and thus fix the ownership-based distribution  rather trivially:  $n_{k}=(N-1)\delta^{0}_{k}+\delta^{1}_{k}$.  Below, for each kind of wealth we shall introduce   what we call  the ``Gentile''  parameter, $\Lambda\in\mathbb{N}$,  which sets  an upper bound on the possession number $k$ as $0\leq k\leq\Lambda$ and  interpolates  boson at $\Lambda=\infty$ and fermion at $\Lambda=1$.   For distinguishable traditional moneys in a `free'  country, the parameter may be set to coincide with the  total number of each kind, \textit{e.g.~}$M$ in (\ref{constraints0}), or to be less by law.  However, electronic forms of wealth can transform to one another. For example, the total amount of deposits at a bank is not fixed due to  the external transfers between accounts  at different banks. The total amount of each  Bitcoin  UTXO  (Unspent Transaction Output)  is not fixed either,  since they can ``combine'' and  ``split'' to other UTXOs~\cite{Bitcoin}.  Thus,  the total number of each species of identical wealth  is not a constant. For this reason and also  a technical reason   later to justify  the approximation of $\ln n_{k}!\simeq n_{k}\ln(n_{k}/e)$, we shall keep $\Lambda$ as an independent key parameter which characterises,  as a matter of principle, boson-like or fermion-like identical wealth.

\textit{Master formula.}---For a  unifying general analysis, we consider  distinguishable and identical wealth together. We call each unit of wealth an \textit{object} and    postulate that  there are $D=d+\brd$  distinct kinds of  {objects}:  $d$ of them are distinguishable and $\brd$ of them are identical.  We label them  by a capital  index, $I=1,2,\cdots, D$, which  decompose into small ones, $I=(i,d+\bri\,)$ where $i=1,2,\cdots, d$ for the distinguishable species and $\bri=1,2,\cdots,\brd$ for the identical species.  An $I$-th kind object has value  $w_{I}\in\mathbb{N}$.  For example, the present-day euro coin series  set ${d=8}$ with ${w_{1}= 1}$,  ${w_{2}=2}$, $\cdots, {w_{8}=200}$  in the unit of cent.  
We then denote a generic ownership over them by  a $D$-dimensional non-negative integer-valued vector,  $\vec{k}=(k_{1},k_{2},\cdots, k_{D})$ of which each component  $k_{I}$  denotes the number of   owned  $I$th-kind objects and is bounded by a cutoff Gentile parameter: $0\leq k_{I}\leq\Lambda_{I}$.  In particular, we set $\Lambda_{I}=\infty$  for bosonic $I$  and  $\Lambda_{I}=1$ for fermionic $I$.   We  let $n_{\vec{k}}$ be the number of the owners with such a ownership~$\vec{k}$. The  total number of owners is then
\be
N\,=\,\sum_{\vec{k}}~n_{\vec{k}}\,\equiv\,\sum_{k_{1}=0}^{\,\Lambda_{1}}\sum_{k_{2}=0}^{\,\Lambda_{2}}\cdots\sum_{k_{D}=0}^{\,\Lambda_{D}}~n_{\vec{k}}\,,
\label{sumN}
\ee
and the  total number of the  $I$th-kind objects is
\be
M_{I}=\sum_{\vec{k}}\,k_{I} n_{\vec{k}\,}\equiv N m_{I}\,.
\label{LI}
\ee
Hereafter,  $\sum_{\vec{k}}$ and $\prod_{\vec{k}}$ are our shorthand notations for the sum  and the product of all $k_{I}$'s from zero to $\Lambda_{I}$'s, as in (\ref{sumN})  above and (\ref{Upsilon}) below. 

On one hand,  as the owners are distinguishable,  the number of    partitions or groupings of the  $N$ owners into  the different  ownerships of $n_{\vec{k}}$'s~(\ref{sumN})  is, generalising (\ref{Upsilon0}),
\be
\Upsilon=\frac{N!}{\,\prod_{\vec{k}}n_{\vec{k}}!\,}\,\equiv\,
\frac{N!}{\prod_{k_{1}=0}^{\Lambda_{1}}\prod_{k_{2}=0}^{\Lambda_{2}}\cdots \prod_{k_{D}=0}^{\Lambda_{D}}\,n_{\vec{k}}!\,}\,.
\label{Upsilon}
\ee
On the other hand for the  partitions of the objects, only the  distinguishable class of objects contributes,  as in (\ref{Phi0}),
\be
\Phi=\prod_{i=1}^{d}\left[\frac{M_{i}!}{\prod_{\vec{k}}\left(k_{i}!\right)^{n_{\vec{k}}}}\right]\,.
\label{Phi}
\ee
For each  partition of owners in $\Upsilon$, any of the  partitions of distinguishable  objects in   $\Phi$ can equally occur. Therefore,  the final, total number of possible outputs for a given set $\{
 n_{\vec{k}}$'s$\}$ is the product,  $\Omega=\Upsilon\times\Phi$.

 We proceed to apply the  variational method to $\ln\Omega$ and aim to acquire the extremal solution  of $n_{\vec{k}}$. While doing so, there are constraints to impose:
\be
\ba{l}
\delta N=\sum_{\vec{k}}~\delta n_{\vec{k}}=0\,,\\
\delta M_{i}=\sum_{\vec{k}}~k_{i}\delta n_{\vec{k}}=0\,,\\
\delta \brM=\sum_{\vec{k}}~\left(\sum_{\bri=1}^{\brd}~w_{\bri}k_{\bri}\right)\delta n_{\vec{k}}=0\,.
\ea
\label{constraints}
\ee
Namely, the total number of owners  and those of distinguishable objects of each kind   are all conserved, as we assume them to be indestructible.  For the identical class of objects, since they may  transform  to other species, we impose that only their total value 
\be
\brM=\sum_{\vec{k}}\left(\textstyle{\sum_{\bri=1}^{\brd}}~w_{\bri\,}k_{\bri}\right)n_{\vec{k}}\equiv N \brm_{w}
\label{brV}
\ee
 is conserved.  To proceed, we employ a well-known approximation for the factorial,   $\ln n_{\vec{k}}!\simeq n_{\vec{k}}\ln (n_{\vec{k}}/e)$, which is valid for large $n_{\vec{k}}$ only.  Our Gentile  cutoff parameter $\Lambda_{I}$ then effectively prevents $n_{\vec{k}}$ from getting too  small, by setting the upper bound on $k_{I}$.  It follows then, from $\delta \ln n_{\vec{k}}!=\delta n_{\vec{k}}\ln n_{\vec{k}}$,  that the variation of $\ln\Omega$  reads 
\be
\delta\ln\Omega=-\sum_{\vec{k}}~\delta n_{\vec{k}}\left[\ln n_{\vec{k}}+\sum_{i=1}^{d}~\ln(k_{i}!)\right]=0\,.
\ee
Around the extremal distribution,  this variation  should vanish, while $\delta n_{\vec{k}}$'s must meet the constraints (\ref{constraints}), otherwise they are arbitrary.  Therefore, only  up to some constants $\alpha, \beta_{i}, \brbeta$, putting   
\be
\alpha\delta N+\left(\sum_{i=1}^{d}\beta_{i}\delta M_{i}\right)+\brbeta\delta\brM-\delta\ln\Omega=0\,,
\ee 
we should have for every $\vec{k}$ without sum,
\be
\ln n_{\vec{k}}+\alpha+\sum_{i=1}^{d}\,\Big[\ln (k_{i}!)+\beta_{i}k_{i}\Big]+\brbeta\sum_{\bri=1}^{\brd}~w_{\bri}k_{\bri}
=0\,.
\ee
This gives  the desired extremal solution,
\be
\ba{ll}
n_{\vec{k}}=N P_{\vec{k}}\,,\qquad&\quad
P_{\vec{k}}=\left[\prod_{i=1}^{d}\,P_{i}(k_{i})\right]\left[\prod_{\bri=1}^{\brd}\,  \brP_{\bri}(k_{\bri})\right]\,,
\ea
\label{Master}
\ee
where $P_{\vec{k}}$ is our  master probability distribution given by the products of $\Lambda$-truncated Poisson and geometric distributions,
\be
\ba{ll}
\dis{P_{i}(k_{i})=\cN_{i~} \frac{e^{-\beta_{i}k_{i}}}{{k_{i}!}}}\,,\quad&\qquad
\dis{\cN_{i}=\frac{1}{\sum_{k_{i}=0}^{\Lambda_{i}}~{e^{-\beta_{i}k_{i}}}/{{k_{i}!}}}}\,,\\
\dis{\brP_{\bri}(k_{\bri})=\cN_{\bri~} e^{-\brbeta w_{\bri} k_{\bri}}}\,,\quad&\qquad
\dis{\cN_{\bri}=\frac{\,1-e^{-\brbeta w_{\bri}}\,~\qquad}{\,1-e^{-(\Lambda_{\bri}+1)\brbeta w_{\bri}}}}\,.
\ea
\label{PG}
\ee
To write this we have solved $\alpha$  in terms of $N$ and the normalisation constants, $\cN_{I}$'s, such that $\sum_{\vec{k}}\,P_{\vec{k}}=1$ and
\be
\ba{l}
\dis{\sum_{\vec{k}}\,{k_{i}P_{\vec{k}}}=\left(1-\cN_{i}\frac{e^{-\beta_{i}\Lambda_{i}}}{\Lambda_{i}!}\right)e^{-\beta_{i}}=m_{i}\,,}\\
\dis{\sum_{\vec{k}}\,{k_{\bri}P_{\vec{k}}}=\frac{1-(\Lambda_{\bri}+1)e^{-\Lambda_{\bri}\brbeta w_{\bri}}+\Lambda_{\bri}e^{-(\Lambda_{\bri}+1)\brbeta w_{\bri}}}{\left(e^{\brbeta w_{\bri}}-1\right)\left[1-e^{-(\Lambda_{\bri}+1)\brbeta w_{\bri}}\right]}\,.}
\ea
\label{kI}
\ee
It remains to determine $\beta_{i}, \brbeta$ from  (\ref{kI}) and (\ref{brV}).   
In particular, when ${\Lambda_{i}=\infty}$,  we get ${e^{-\beta_{i}}= m_{i}}$ and  a  precise Poisson distribution  holds with $\cN_{i}=e^{-m_{i}}$. On the other hand, when ${\brd=1}$ and ${\Lambda_{\bri}=\infty}$ or ${\Lambda_{\bri}=1}$, we obtain ${e^{-\brbeta w_{\bri}}=\frac{m_{\bri}}{1\pm m_{\bri}}}$ and recover  the  Bose--Einstein or Fermi--Dirac distributions  having an exponential tail,
\be
m_{\bri}\,=\,\sum_{\vec{k}}\,{k_{\bri}P_{\vec{k}}}\,=\,\frac{1}{e^{\brbeta w_{\bri}}\mp1}\,,
\ee
which quantify  the `popularity' (or inverse `rarity' \textit{c.f.}~\cite{rarity}) of the  digital  wealth. 
As  the geometric distribution  is essentially the exponential   Boltzmann--Gibbs law, we may identify $\brbeta$ as the inverse ``temperature'',  see also \cite{BGL}.

The distribution of the total value follows
\be
\cP(v)=\sum_{\vec{k}}~\scalebox{1.1}{$\delta$}_{\vec{w}{\cdot\vec{k}}}^{\displaystyle{v}}~P_{\vec{k}}\,,
\label{weightedP}
\ee
where $\scalebox{1.1}{$\delta$}_{\vec{w}{\cdot\vec{k}}}^{\displaystyle{v}}$ is the Kronecker-delta with ${\vec{w}{\cdot\vec{k}}}=\sum_{I=1}^{D}w_{I}k_{I}$ amounting to a total value~$v$.  Essentially (\ref{weightedP}) is a weighted convolution   whose generating function  reads for $\Lambda_{i}=\infty$,
\be
\ba{lll}
Z(q)&=&\dis{\sum_{v=0}^{\infty}~\cP(v)q^{v}\,=\,\sum_{\vec{k}}~P_{\vec{k}\,} q^{\vec{w}\cdot{\vec{k}}}=\left[\prod_{i=1}^{d}~e^{m_{i}\left(q^{w_{i}}-1\right)}\right]}\\
\multicolumn{3}{l}{\dis{\,\qquad\times{\left[\prod_{\bri=1}^{\brd}\left(\frac{e^{\brbeta w_{\bri}}-1}{e^{\brbeta w_{\bri}}-q^{w_{\bri}}}\right)
\left(\frac{e^{(\Lambda_{\bri}+1)\brbeta w_{\bri}}-q^{(\Lambda_{\bri}+1)w_{\bri}}}{e^{(\Lambda_{\bri}+1)\brbeta w_{\bri}}-1}\right)
\right]\,.}}}
\ea
\label{MasterZ}
\ee
While the truncated Poisson distribution $P_{i}(k_{i})$~(\ref{PG}) with a finite cutoff $\Lambda_{i}$ can be applicable to rare valuable  items that are not necessarily hard cash, henceforth, for simplicity,  we set    ${\Lambda_{i}=\infty}$ (distinguishable) and ${\Lambda_{\bri}=\infty}$ (bosonic) or ${\Lambda_{\bri}=1}$ (fermionic).\footnote{The geometric distribution~$\brP_{\bri}(k_{\bri})$~with other finite values of  $\Lambda_{\bri}$ appears applicable to  some  Ethereum’s flexible token standard (ERC-1155)~\cite{ERC1155}.}   The Poisson and the bosonic/fermionic geometric distributions
\be
\ba{ll}
P_{\mP}(m,k)=e^{-m\,}\frac{m^{k}}{k!}\,,\quad&\quad
\brP_{\mb}(m,k)=
\frac{1}{1+m}\Big({\frac{m}{1+m}}\Big)^{k}\,,\\
\multicolumn{2}{l}{~\brP_{\mf}(m,k)=(1-m)\delta^{0}_{k}+m\delta^{1}_{k}=
(1{-m})\Big({\frac{m}{1-m}}\Big)^{k}\,,}
\ea
\label{PbfG}
\ee
are then the  elemental `atomic' distributions in (\ref{PG}). Here, ${m>0}$ is the  mean value in each distribution. For the fermionic distribution, it should be less than one, such as $m=1/N$. Further, the variance is $m$ or   
$m(1\pm m)$ for the distinguishable or bosonic/fermionic cases.  In the vanishing limit $m\rightarrow 0$, they all reduce to a Kronecker-delta distribution: $P_{\mP}(0,k)=\brP_{\mbf}(0,k)=\delta_{k}^{0}$.   

\textit{Poisson versus Geometric}.---As relevant to both financial assets and cryptocurrencies,  here we make  various  comparisons between  $P_{\mP}(m,k)$  and $\brP_{\mb}(m,k)$ allowing  arbitrary $m>0$ and unrestricted $k=0,1,2,\cdots,\infty$.

\begin{figure}[h]
\includegraphics[width=80mm]{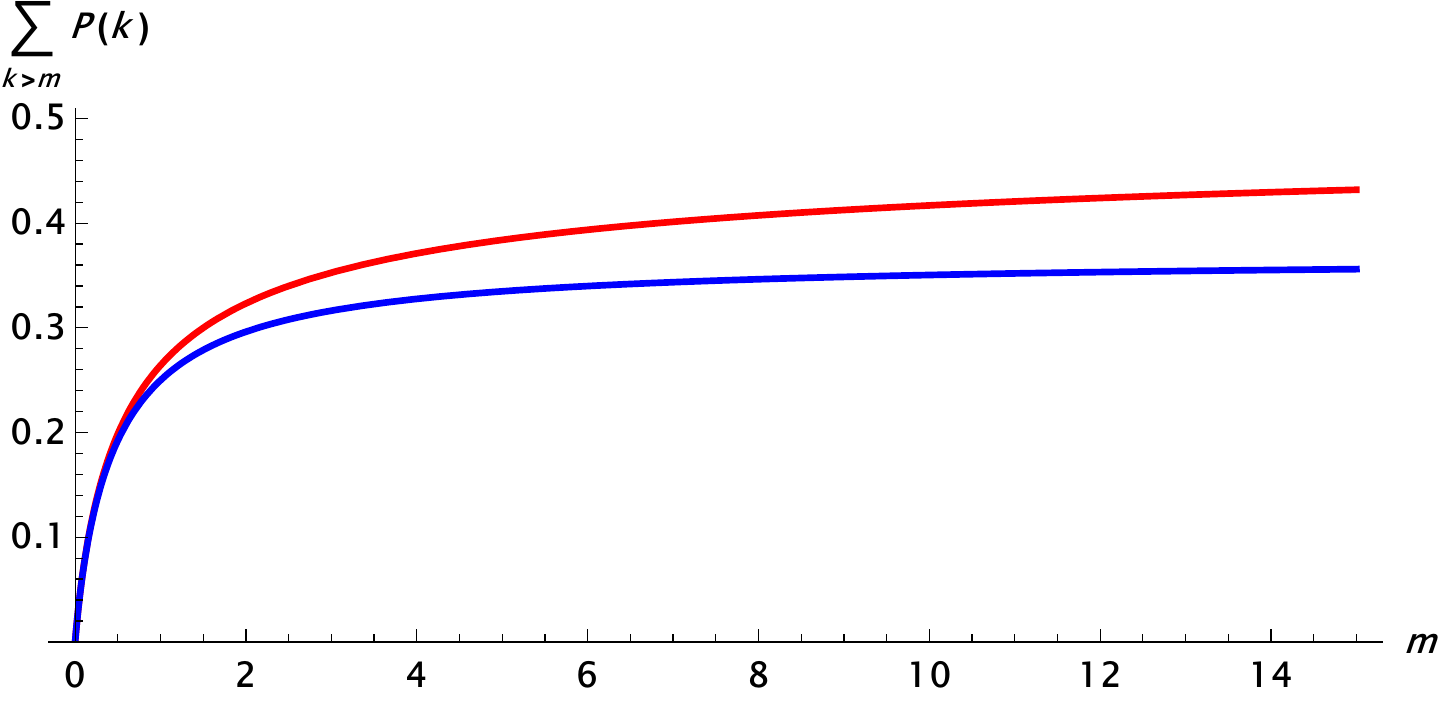}
\caption{The  probability to own more than  mean  value $m$:   $\sum_{k>m}\,P_{\mP}(m,k)$ (Poisson for distinguishable wealth, red)  vs. $\sum_{k>m}\,\brP_{\mb}(m,k)$ (geometric for identical wealth, blue),  with varying mean value $m$ (horizontal axis). The former   is always larger than the latter. They converge to ${1/2}$ and ${e^{-1}\simeq 0.367879}$   in the large $m$ limit (\ref{halfsumlimit}). \label{FIGupper} }
\end{figure}

While  $\brP_{\mb}(m,k)$  is a monotonically decreasing function in $k$,  
from Stirling's formula, $\ln k!\simeq k\ln k-k+\ln\sqrt{2\pi k}$, 
$P_{\mP}(m,k)$   assumes the maximal value, 
\be
{\rm{Max}}\big[P_{\mP}(m,k)\big]\simeq 1/\sqrt{2\pi m}~~~\mbox{at}~~k\simeq m\,.
\label{MaximalP}
\ee
That is to say,   the Poisson distribution is on-peak   for the  owners of  the  averaged   wealth $m=M/N$, namely the `middle class'.  Further, the ratio of the two distributions
\be
{\brP_{\mb}(m,k)}/{P_{\mP}(m,k)}={e^{m}k!}/{(m+1)^{k+1}}
\ee
shows that the  geometric distribution  has a thicker tail  than Poisson  one for ${k>>m}$. Yet,  complementary to this, an inequality holds:
\be
\sum_{k>m}~\brP_{\mb}(m,k)~<~\sum_{k>m}~P_{\mP}(m,k)\,,
\ee
which implies that the probability for $k>m$  is larger  in the Poisson distribution compared to the geometric one, see FIG.\,\ref{FIGupper}. In fact, in the large $m$ limit, we have~\cite{NIST}
\be
\ba{ll}
\dis{\lim_{m\rightarrow\infty}\!\sum_{k=m{+1}}^{\infty}\!\!\!P_{\mP}(m,k)=\frac{1}{2}\,,}~&~
\dis{\lim_{m\rightarrow\infty}\!\sum_{k=m{+1}}^{\infty}\!\!\!\brP_{\mb}(m,k)=e^{-1}\,.}
\ea
\label{halfsumlimit}
\ee
Thus,   $50\%$ or about $37\%$ of the holders have more than the mean value in the Poisson or geometric distribution.

We compare Shannon entropy,  $S=\sum_{k}\,-P(k)\ln P(k)$. 
Since both $P(k)$ and $-\ln P(k)$ are non-negative, the entropy is bounded $S\geq 0$. The saturation occurs when everyone has the equal amount of wealth~\textit{i.e.}~the average value $m$ implying  $P(k)=\delta^{m}_{k}$, \textit{i.e.~}either $P(k)=0$ or $\ln P(k)=0$.  For the Poisson and geometric distributions, this  happens only in the vanishing limit $m\rightarrow 0$.  For a given  arbitrary value of  $m$, it is famously  the geometric distribution~$\brP_{\mb}(m,k)$ that sets the entropy maximal, 
\be
\bar{S}_{\mb}(m)=(m+1)\ln(m+1)-m\ln m\,.
\label{brSmb}
\ee
The entropy of the Poisson distribution~$P_{\mP}(m,k)$~\cite{EntropyPoisson},
\be
S_{\mP}(m)=\frac{1}{2}\ln(2\pi em)-\frac{1}{12m}+O(m^{-2})
\ee
is then roughly  half of the maximum~(\ref{brSmb}) for large $m$.

\begin{figure}[h]
\includegraphics[width=80mm]{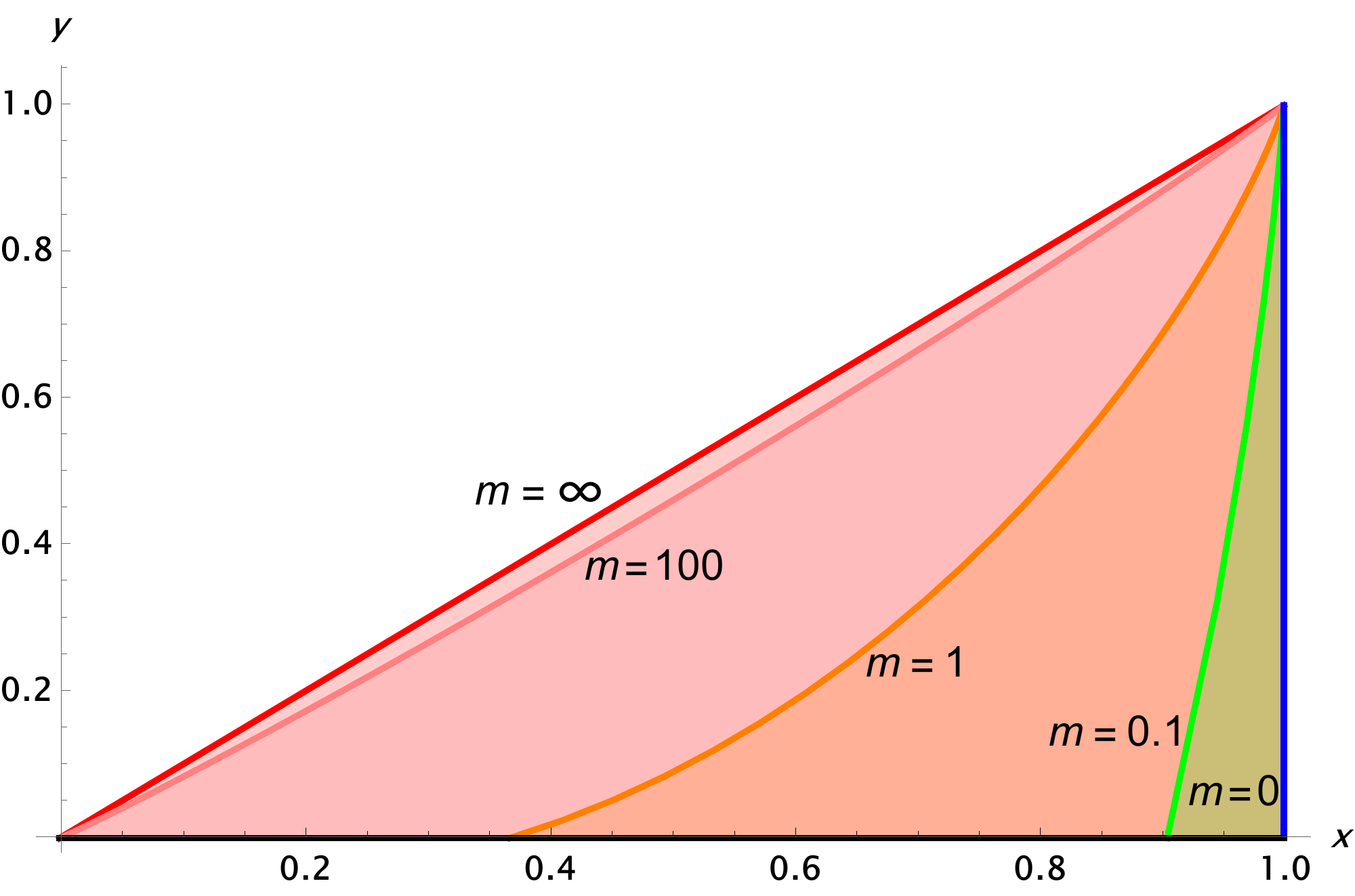}
\caption{Lorenz curves of the Poisson distribution $P_{\mP}(m,k)$ for distinguishable wealth. \textit{i)} ${m=\infty}$, ${\Gini_{\mP}=0}$ ($45$-degree line of perfect equality), 
 \textit{ii)} ${m=100}$, ${\Gini_{\mP}\simeq 0.056}$, \textit{iii)} ${m=1}$,  ${\Gini_{\mP}\simeq 0.52}$,  \textit{iv)} ${m=0.1}$,  ${\Gini_{\mP}\simeq 0.91}$, and \textit{v)} ${m=0}$,   ${\Gini_{\mP}=1}$ as ${y=\delta^{0}_{x}\,}$.  Each curve  includes ${y=0}$ for an interval $0\leq x\leq e^{-m}$.   Only when $m\approx  0.35$,   ``$ 80/20$ rule''  holds. \label{FIGLP} }

\vspace{5pt}

\includegraphics[width=80mm]{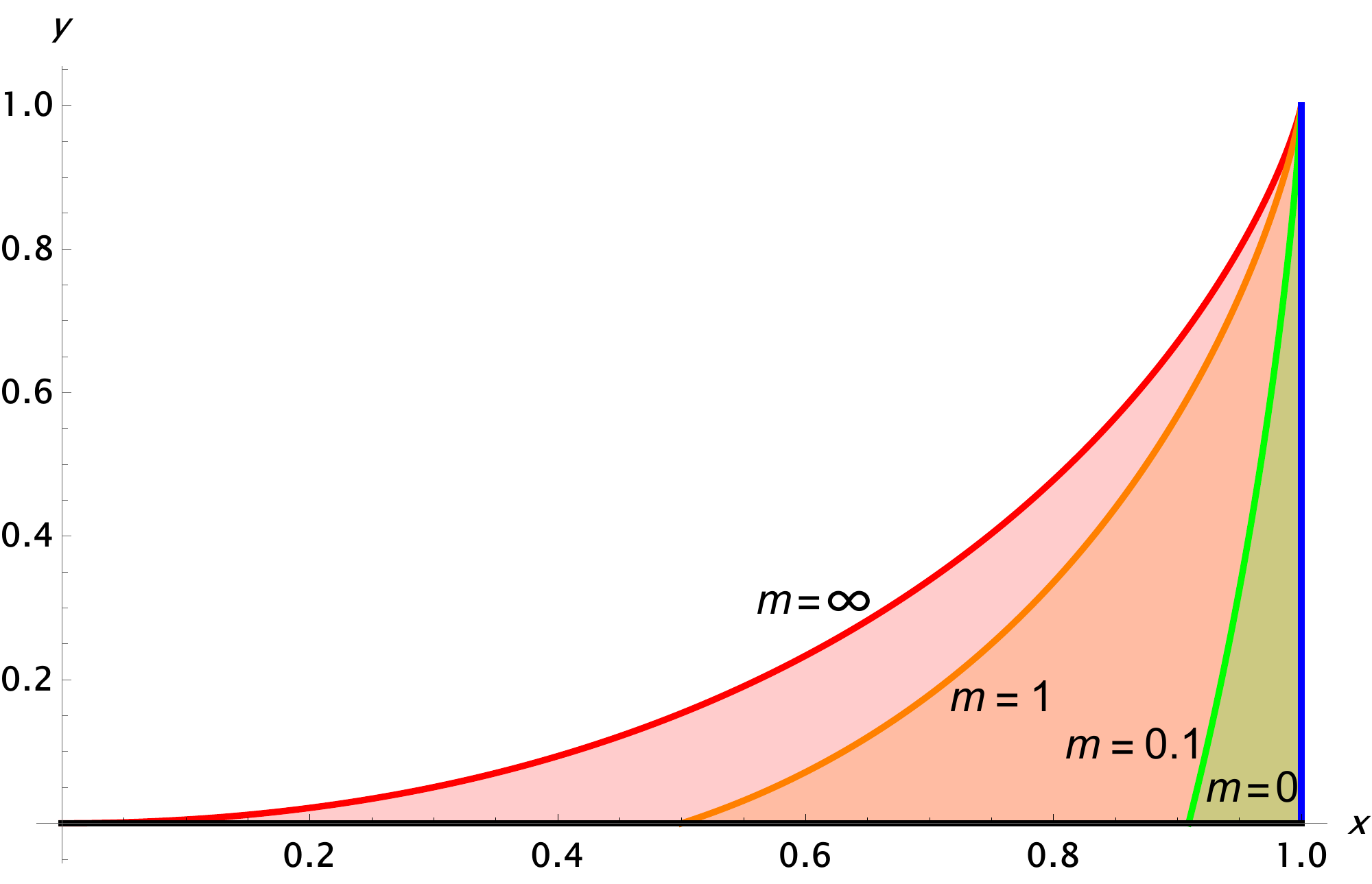}
\caption{Lorenz curves of the geometric distribution $\brP_{\mb}(m,k)$ for identical wealth.  \textit{i)} ${m=\infty}$, ${\Gini_{\mb}=\frac{1}{2}}$ as saturated by $y=x+(1-x)\ln(1-x)$~\cite{BGL2}, \textit{ii)} ${m=1}$,  ${\Gini_{\mb}\simeq 0.68}$,  \textit{iii)} ${m=0.1}$,  ${\Gini_{\mb}\simeq 0.93}$, and \textit{iv)} ${m=0}$,   ${\Gini_{\mb}=1}$  as ${y=\delta^{0}_{x}}$. Each curve includes ${y=0}$ for an interval $0\leq x\leq\frac{1}{m+1}$. 
From (\ref{LorenzG}), only when $m\approx  0.47$,    ``$ 80/20$ rule (aka  Pareto principle)''  holds. 
\label{FIGLG}}
\end{figure}

We draw the   Lorenz curves of $P_{\mP}(m,k)$  and $\brP_{\mb}(m,k)$ as FIG.\,\ref{FIGLP} and FIG.\,\ref{FIGLG},  by setting $x=\sum_{j=0}^{k}\,P(j)$ and $y=\frac{1}{m}\sum_{j=0}^{k}\,j\,P(j)$.
  Since ${P(0)\neq 0}$  in both cases, the curves should include an interval ${0\leq x\leq P(0)}$ for trivial ${y=0}$. While we depict  the Lorenz curve of $P_{\mP}(m,k)$  numerically,  for the geometric distribution~$\brP_{\mb}(m,k)$,  we solve   for $k$  in terms of $x$, 
\be
k+1=-\,\frac{\ln(1-x)}{\ln(1+1/m)}\,,
\label{kp1}
\ee
and obtain an analytic expression of the Lorenz  curve:
\be
y(x)=\left\{\ba{cll}
\dis{x+\frac{(1-x)\ln(1-x)}{m\ln(1+{1}/{m})}}&~\mbox{for}~& \dis{\frac{1}{m+1}\leq x < 1}\\
0&~\mbox{for}~&\dis{ 0\leq x\leq \frac{1}{m+1}}\ea\right.\,,
\label{LorenzG}
\ee
of which the large $m$ limit is known~\cite{BGL2}.

Lastly, we  compute the Gini coefficient defined by
\be
\!\ba{lrl}
\Gini[m]&:=&
\sum_{k=0}^{\Lambda}\sum_{k^{\prime}=0}^{\Lambda}~\frac{\left| k-k^{\prime}\right|}{2m}P(k)P(k^{\prime})\\
{}&=&
1+\frac{1}{m}\sum_{k=0}^{\Lambda}\,P(k)\left[k P(k)
-2\sum_{k^{\prime}=0}^{k}\,k^{\prime}P(k^{\prime})\right]\,.
\ea
\label{GiniSUM}
\ee
For $P_{\mP}(m,k)$,  from $\frac{1}{(k!)^{2}}=\frac{1}{\pi (2k)!}\scalebox{1.2}{$\int_{0}^{\pi}$}d\theta\,(2\cos\theta)^{2k}$,  we get \textit{c.f.~}\cite{Gini1}
\be
\Gini_{\mP}[m]=\dis{\frac{1}{\pi}\int_{0}^{\pi}\!d\theta~e^{-2m(1-\cos\theta)}(1+\cos\theta)}\,.
\label{GiniP}
\ee   
For  $\brP_{\mb}(m,k)$ and additionally  $\brP_{\mf}(m,k)$, we have\footnote{
Alternative to (\ref{GiniSUM}), we may compute the Gini coefficient through an integral of the Lorenz curve~(\ref{LorenzG}),
\[
\Gini_{\mb}^{\prime}[m]=\left(\frac{m}{m+1}\right)^{2}\left(\frac{1}{2m\ln(1+1/m)}+\frac{1}{m}+\frac{1}{m^{2}}\right)\,,
\label{GiniG2}
\]
which differs from $\Gini_{\mb}[m]$ in (\ref{Ginibf})  by at most $2.4\%$ at   $m\simeq0.53$.}
\be
\ba{ll}
\Gini_{\mb}[m]=\frac{1+m}{1+2m}\,,\quad&\qquad
\Gini_{\mf}[m]=1-m\,.
\ea
\label{Ginibf}
\ee
We note then
\be
\ba{rcl}
\Gini_{\mP}[m]\,<\,\Gini_{\mb}[m]&\quad\mbox{for}&\mbox{\,arbitrary\,}~m>0~~\mbox{and}\\
\Gini_{\mf}[m]\,<\,\Gini_{\mP}[m]\,<\,\Gini_{\mb}[m]&\quad\mbox{for}\quad&\quad 0< m<1\,.
\ea
\ee
Especially in the  large $m$ limit,  we get ${{\Gini_{\mP}}[\infty]=0}$ (the perfect equality)  and  ${{\Gini_{\mb}}[\infty]=\frac{1}{2}}$.  In the opposite vanishing limit, the Gini coefficients are all unity, ${\Gini_{\mP,\mb,\mf}[0]=1}$, hence economically most unequal. Though the fermionic   Gini coefficient $\Gini_{\mf}[m]=1-m$ can be close to unity as $m=1/N<<1$,  due to the severe restriction of the possession, \textit{i.e.~}${k=0,1}$, it is the smallest among the three.

\textit{More than one bank.}---We now consider the deposits of savings accounts at more than one bank which allow  external transfers and  adopt the same minimal unit of currency.  That  corresponds  to the equal-weighted convolution~(\ref{weightedP}) of the geometric distributions:  with ${w_{\bri}\equiv 1}$,
\be
\ba{l}
\dis{\brP_{\brd\,}(m,k)=\frac{(\brd+k-1)!}{(\brd-1)!\,k!}\left(\frac{\brd}{m+\brd}\right)^{\brd}\left(\frac{m}{m+\brd}\right)^{k}\,,}\\
\dis{
\bar{Z}_{\brd\,}(m,q)=\sum_{k=0}^{\infty}~\brP_{\brd\,}(m,k)q^{k}
=\left[\frac{\brd}{\brd-m(q-1)}\right]^{\brd}
\,,}
\ea
\label{brPbrd}
\ee
where $\brd$ is the number of the banks.  Remarkably,\footnote{In contrast, 
rather natural  from the very  distinguishability, the  equal-weighted convolution  of the   Poisson distributions is \textit{closed}:
\[
\sum_{l=0}^{k}~ P_{\mP}(m_{1},l)P_{\mP}(m_{2},k-l)\,=\,P_{\mP}(m_{1}+m_{2},k)\,.
\]} for $\brd\geq 2$, $\brP_{\brd\,}(m,k)$ is no longer a monotonically decreasing function in $k$. It assumes the maximal value, 
\be
\textstyle{
{\rm{Max}}\big[P_{\brd\,}(m,k)\big]\simeq \frac{1}{\sqrt{2\pi m\left(1-\frac{1}{\brd}\right)\left(1+\frac{m}{\brd}\right)}}~~\mbox{at}\,~{k^{\star}\simeq \left(1-\frac{1}{\brd}\right) m\,.}}
\label{MaximalB}
\ee
The fact  ${k^{\star}<m}$ implies that $\brP_{\brd}(m,k)$ is a more unequal distribution compared to the Poisson one~$P_{\mP}(m,k)$~(\ref{MaximalP}). Nonetheless, in the large $\brd$ limit, $\brP_{\brd\,}(m,k)$, $\brZ_{\brd\,}(m,q)$, and the maximum~(\ref{MaximalB}) all  reduce to  those of the Poisson distribution or (\ref{MaximalP}),
\be
\ba{ll}
\dis{\lim_{\brd\rightarrow\infty}\,\brP_{\brd\,}(m,k)=
e^{-m\,}\frac{m^{k}}{k!}}\,,\quad&\quad
\dis{\lim_{\brd\rightarrow\infty}\,\bar{Z}_{\brd\,}(m,q)=e^{m(q-1)}}\,.
\ea
\label{limit}
\ee
An intuitive explanation is as follows. When the number of the banks is infinite,  each bank has most  likely zero or only one unit of the deposits.   The identical wealth then effectively becomes  distinguishable by the distinct banks.  In this way, $\brP_{\brd\,}(m,k)$ interpolates the geometric and the Poisson distributions, or FIG.\,\ref{FIGLP} and FIG.\,\ref{FIGLG}. More banks there are, smaller the Gini coefficient is.

\textit{Boson-like Bitcoin.}---As a cryptocurrency, Bitcoin~\cite{Bitcoin}  belongs to the  identical class of wealth.  Although  each UTXO has its unique cryptographic hash, it generates   insignificant ignorable information.   UTXOs  of a common value are identical, while those of  different values are distinguishable, \textit{c.f.~}\cite{Kahn,Milne}.   The value of every  UTXO is   discretised   in a minimal unit called `satoshi'. In this unit, we have $w_{\bri}\equiv\bri$ where $\bri$ runs from one to $\brd=2.1\times 10^{15}$ which is the  hard cap encoded in Bitcoin’s source code. For each UTXO worthy of  $\bri$ satoshi, the  ownership-based distribution and the  expected number  are from (\ref{PG}) given by  geometric and Bose--Einstein distribution respectively, 
\be
\ba{ll}
\dis{\brP_{\bri}(k_{\bri})=\left(1-e^{-\bri \brbeta}\right)e^{-\bri\brbeta k_{\bri}}}\,,\quad&\quad
\dis{\sum_{k_{\bri}=0}^{\infty}~k_{\bri}\brP_{\bri}(k_{\bri})=\frac{1}{e^{\bri\brbeta}-1}\,.}
\ea
\label{Bitcoin}
\ee
The generating function of the total value~(\ref{MasterZ}) is then
\be
\dis{Z(q)=\prod_{\bri=1}^{\brd}\frac{1-e^{-\bri\brbeta}}{1-\big(e^{-\brbeta}q\big)^{\bri\,}}=\sum_{v=0}^{\infty}~\cP(v)q^{v}}\,,
\ee
and thus, for $v\leq\brd$ the total-value-based distribution is 
\be
\ba{ll}
\dis{
\cP(v)=\cP(0)\mathfrak{P}(v) e^{-v\brbeta}\,,}\qquad&\quad
\dis{\cP(0)=\prod_{\bri=1}^{\brd}\left({1-e^{-\bri\brbeta}}\right)\,,}
\ea
\ee
where $\mathfrak{P}(v)$ is    the  number-theory  \textit{partition} of the non-negative integer $v$, which   appears here  since the UTXO values are equally spaced \textit{i.e.~}$w_{\bri}=\bri$, as is the case with a simple harmonic quantum oscillator.

We need to  determine $\brbeta$ in terms of the mean  total  value, \textit{i.e.~}$\brm_{w}=\bar{M}_{w}/N$~(\ref{brV}), 
\be
\dis{\sum_{s=0}^{\infty}~s\cP(s)=\left.q\partial_{q} Z(q)\right|_{q=1}=
\sum_{\bri=1}^{\brd}~\frac{\bri}{e^{\bri \brbeta}-1}=\brm_{w}}\,.
\label{brbetamw}
\ee
Practically putting ${\brd=\infty}$, we approximate the above sum  by a semi-infinite integral,
\be
\sum_{\bri=1}^{\brd}~\frac{\bri}{e^{\bri \brbeta}-1}\,\simeq\, \brbeta^{-2}
 \int_{0}^{\infty}\rd x~\frac{x}{e^{x}-1}=\frac{\pi^{2}}{6\brbeta^{2}}\,,
\label{sumint}
\ee
and fix  $\brbeta$,
\be
\brbeta\simeq\frac{\pi}{\sqrt{6\brm_{w}}}\,.
\label{brbeta}
\ee
Further, from  the Hardy--Ramanujan formula of the partition, we obtain for large enough $v$, 
\be
\frac{\cP(v)}{\cP(0)}\simeq\frac{1}{4v\sqrt{3}}\,e^{\pi\sqrt{2v/3}\,-\,v\brbeta}\,,
\label{HardyRamanujan}
\ee
such that its maximum
\be
{\rm{Max}}\left[\frac{\cP(v)}{\cP(0)}\right]\simeq \frac{\sqrt{3}\brbeta^{2}}{2\pi^{2}}\,e^{{(\pi^{2}/6)}\brbeta^{-1}}
\label{MaximalcP}
\ee
is positioned at ${v^{\star}}$ which is  smaller than the mean value,
\be
v^{\star}\simeq \frac{\pi^{2}}{6\brbeta^{2}}\left(\frac{1+\sqrt{1-24\brbeta/\pi^{2}}}{2}\right)^{2}\,<\,\brm_{w}\,=\,\frac{\pi^{2}}{6\brbeta^{2}}\,.
\ee
This  inequality implies that, despite  the large $\brd$  limit which we have tactically assumed, in contrast to the many bank limit~(\ref{limit}),   the Bitcoin distribution with ${w_{\bri}=\bri}$  is still more unequal  than the Poisson one~(\ref{MaximalP}): $\cP(v)$~(\ref{HardyRamanujan}) has thicker tail than $P_{\mP}(m,k)\sim(me/k)^{k}$.

According to \cite{BitcoinDATA}, as of 2022,  the total number of addresses reads $N\sim 10^{9}$,  and the total value of all the UTXOs is roughly $\bar{M}_{w}\sim 10^{15}$~satoshi. We then  estimate  $\brm_{w}\sim 10^6$ and, from (\ref{brbeta}), $\brbeta\sim 10^{-3}$, the smallness of which  justifies our  integral approximation~(\ref{sumint}).\footnote{For ${\brbeta=10^{-3}}$  and  $\brd\geq 10^4$,  the error of (\ref{sumint}) is less than $0.1\%$. }

\textit{Discussion.}---To conclude,  traditional tangible moneys are distinguishable; yet  financial assets and    cryptocurrencies  are all identical.  The usage of the boson-like  wealth   results in more unequal geometric-type distribution  compared to the Poisson-type distribution of the distinguishable  wealth.  While so, aggregating different kinds of wealth    leads  to a weighted convolution.  In particular, the existence of more than one bank softens the economic inequality of the  geometric distribution by  a monopolistic   bank.   {Similar to (\ref{limit}) which is for  bosonic  geometric distributions,  the   equal-weighted-convolution   of fermionic  geometric  distributions~(\ref{PbfG})  also  converges to a Poisson distribution  in the  large limit of  total amount $\bar{M}$ with  fixed mean value $m=\bar{M}/N$\,: the (binomial) convolution
\be
\brP_{\bar{M}}(m,k)=\frac{\bar{M}!}{(\bar{M}-k)!k!}\left(1-\frac{1}{N}\right)^{\bar{M}-k}\left(\frac{1}{N}\right)^{k}
\ee
converges to a Poisson distribution,
\be
\dis{\lim_{\bar{M}\rightarrow\infty}\,\brP_{\bar{M}}(m,k)=
e^{-m\,}\frac{m^{k}}{k!}}\,.
\label{limit2}
\ee
This provides an alternative derivation of the Poisson distribution of distinguishable objects. Even though hard cashes are distinguishable, each of them   is unique and thus its distribution should coincide with that of NFT, \textit{i.e.~}the fermionic geometric distribution~(\ref{PbfG}). After considering multiple of them of the same value, through the equal-weighted-convolution, the Poisson distribution emerges consistently  out of  the bosonic  as well as  fermionic geometric distributions, (\ref{limit}) and (\ref{limit2}).

}

The distribution of Bitcoin  is given by the number-theory partition.   For completeness, the convolution of a geometric and a Poisson distribution, as for hard cash and savings account, reads
\be
\ba{lrl}
\hat{P}(m,\brm,k)&:=&\dis{\sum_{j=0}^{k}\, P(m,j)\brP(\brm,k-j)}\\
{}&=&\dis{\frac{e^{-m}}{\brm+1}\left(\frac{\brm}{\brm+1}\right)^{\!k\,}\sum_{j=0}^{k}\frac{1}{j!}\big(m+m/\brm\big)^{j}}\,,
\ea
\ee
which carries a power-law tail $\,\frac{e^{m/\brm}}{\brm+1}\big(\frac{\brm}{\brm+1}\big)^{k}$ for large $k$.

Putting $w_{\bri}=1$ and $w_{\bri}=-1$ separately for a pair of  $\brP_{\brd}(m,k)$'s~(\ref{brPbrd}), we can further aggregate deposit and debt: for net balance $a\in\mathbb{Z}$, we have
\be
\cP_{\brd}(m_{1},m_{2},a):=\sum_{k_{1}=0}^{\infty}\sum_{k_{2}=0}^{\infty}~\delta^{a}_{k_{1}-k_{2}}\cP_{\brd}(m_{1},k_{1})\cP_{\brd}(m_{2},k_{2})\,,
\ee
where ${m_{1}\geq 0}$ and ${m_{2}\geq 0}$ are the mean values of deposit and debt respectively.  
In particular, for ${\brd=1}$ we get
\be
\cP_{\brd=1}(m_{1},m_{2},a)=\left\{\ba{lll}
\frac{1}{m_{1}+m_{2}+1}\left(\frac{m_{1}}{m_{1}+1}\right)^{a}&\mbox{~for~}&a\geq 0\\
\frac{1}{m_{1}+m_{2}+1}\left(\frac{m_{2}}{m_{2}+1}\right)^{\left|a\right|}&\mbox{~for~}&a< 0\,.
\ea\right.
\ee

\textit{A priori}, the Poisson and  geometric distributions~(\ref{PbfG}) depend on the  mean  `number'  ${m=M/N}$  (dimensionless),  rather than any   `value' (``dimensionful'').  Therefore,  any adjustment  of the minimal unit, \textit{e.g.~}demolishing cents and keeping euros only,  can change the number $M$ and  affect the distributions.

 It would be of interest to investigate any   phase transition for the master distribution~(\ref{Master}) through the changes of variables, even if $N$ is finite~\textit{c.f.~}\cite{7616}.   As Bitcoin is boson-like, one may wonder about  Bose--Einstein condensation especially  to the minimal ${\bri=1}$ UTXO. For this, we consider its popularity normalised by  the mean total  value~(\ref{brbetamw}), or the ratio $\frac{1}{e^{\brbeta}-1}/\big[\sum_{\bri=1}^{\infty}~\frac{\bri}{e^{\bri \brbeta}-1}\big]$. This quantity  increases monotonically from zero at $\brbeta=0$ and converges to one as $\brbeta$ grows. In particular, when $\brbeta\geq 3$, it becomes  greater than $0.9$.  This ``low temperature'' might be attainable if   Bitcoin  gets ever extremely popular:  (somewhat unrealistically) large $N$ with  $\bar{M}_{w}$ bounded by  the hard cap.

We have restricted our work to be  theoretical. Yet, the resulting distributions including   FIG.\,\ref{FIGLP} and FIG.\,\ref{FIGLG}  appear  consistent  with real data,  for example  \cite{Sweden,Norway,BGL3}.  Besides,   the (truncated) Poisson-type distribution (\ref{PG}) can be 
applied not only to tangible moneys, but also to various objects, including  citations of research papers~\cite{citation}.   

 Taking into account the  individual  differences of owners, or other extra factors,   may weaken  the assumed   `randomness'.   Even  so,    we expect that the difference of    inequality  in  distributions   persists depending on the class of wealth, distinguishable or identical.  We call for thorough verifications with wide  applications.

Lastly, while we have borrowed the notion of indistinguishability from  particle \& statistical  physics for the description of   financial wealth,  namely \textit{econophysics}~\cite{Stanley1,Stanley2,Nature2013},   our results like (\ref{limit}) may help to understand  how macroscopic objects formed by many identical particles appear distinguishable, \textit{i.e.~}through the generation of large degeneracy of  quantum states.\\

{\textit{Acknowledgments.}}---We wish to   thank  Marc Jourdan, Chunghyoung  Lee, Sukgeun Lee, 
  Hocheol Lee,  and \href{https://glassnode.com}{Glassnode Support Team} for helpful communications. This work is supported by Basic Science Research Program through the National Research Foundation of Korea (NRF)  Grants, NRF-2016R1D1A1B01015196 and  NRF-2020R1A6A1A03047877 (Center for Quantum Spacetime). 
\hfill

\end{document}